\documentclass[utf8]{frontiersinFPHY_FAMS} 

\usepackage{tabularx}
\usepackage{booktabs}
\usepackage{makecell}

\usepackage{amsmath, amssymb, amsfonts,mathtools} 
\usepackage{units} 
\usepackage{eurosym}
\usepackage{csquotes} 
\usepackage[nocomma, short]{optidef}

\usepackage{url}

\usepackage{url,hyperref,lineno,microtype,subcaption}
\usepackage[onehalfspacing]{setspace}
\usepackage[T1]{fontenc} 
\usepackage{tikz}
\usetikzlibrary{arrows.meta}

\usepackage[dvipsnames]{xcolor}

\linenumbers

\def\keyFont{\fontsize{8}{11}\helveticabold }
\def\firstAuthorLast{Au-Yeung {et~al.}} 
\def\Authors{Rhonda Au-Yeung\,$^{1,*,\dagger}$, Nicholas Chancellor\,$^{2,\dagger}$, and Pascal Halffmann\,$^{3,*,\dagger}$}
\def\Address{$^{1}$Department of Physics, University of Strathclyde, Glasgow G4 0NG, United Kingdom,\\
$^{2}$Department of Physics, Durham University, Durham DH1 3LE, United Kingdom,\\
$^{3}$ Department of Financial Mathematics, Fraunhofer Institute for Industrial Mathematics, Am Fraunhofer Platz 1, 67663 Kaiserslautern, Germany}
\def\corrAuthor{Corresponding Authors}

\def\corrEmail{rhonda.au-yeung@strath.ac.uk\\ pascal.halffmann@itwm.fraunhofer.de}

\begin{document}
	\nolinenumbers
\onecolumn
\firstpage{1}

\title {NP-hard but no longer hard to solve? Using quantum computing to tackle optimization problems} 

\author[\firstAuthorLast ]{\Authors} 
\address{} 
\correspondance{} 

\extraAuth{}

\maketitle

\begin{abstract}
In the last decade, public and industrial research funding has moved quantum computing from the early promises of Shor's algorithm through experiments to the era of noisy intermediate scale quantum devices (NISQ) for solving real-world problems. It is likely that quantum methods can efficiently solve certain (NP-)hard optimization problems where classical approaches fail. In our perspective, we examine the field of \emph{quantum optimization} where we solve optimisation problems using quantum computers. We demonstrate this through a proper use case and discuss the current quality of quantum computers, their solver capabilities, and benchmarking difficulties. Although we show a proof-of-concept rather than a full benchmark, we use the results to emphasize the importance of using appropriate metrics when comparing quantum and classical methods. We conclude with discussion on some recent quantum optimization breakthroughs and the current status and future directions.
\tiny
\keyFont{ \section{Keywords:} Quantum Computing (QC), Quantum Optimization, Problem Modeling, Optimization Algorithms, Perspective} 
\end{abstract}

ORCID: R. A. (0000-0002-0082-5382), N. C. (0000-0002-1293-0761), P. H. (0000-0002-3462-4941)

\section{Introduction}\label{sec:intro}

In November 1994, Peter Shor presented his pioneering algorithm at the IEEE Symposium on Foundations of Computer Science in Santa Fe, NM \cite{Shor1994}. Without a doubt, Shor's algorithm threatened the underpinnings of public key cryptography, putting quantum computing on the proverbial map and forcing people outside the academic community to pay attention \cite{Witt2022}. Decades later, it remains the standard-bearer of quantum algorithms. Almost as important, Shor's algorithm has currently led to investments of billions of (US) dollars worldwide into quantum computing technologies\footnote{e.g. the German government has invested two billion euros in quantum computing \cite{Germany2021}
}. This drastic increase of funding has seen a spike in quantum computing research at public and private research institutes, R\&D departments in the commercial sector, as well as the founding of new startups \cite{Brooks2019,Gibney2019,MacQuarrie2020}. As of 15 Dec 2022, 10,416 quantum physics articles have been published on the arXiv preprint server where nearly half these papers involve quantum computing. In tandem with the volume of publications, quantum computers have evolved from experiments to an era of noisy intermediate scale quantum (NISQ) devices \cite{Callison2022,Preskill2018} where they play an increasingly popular role in real-world problems \cite{Bova2021}.

One exciting application area is optimization. In business, time is money. Everything can be reduced to an optimization problem. Every company decision is weighed against resources, financial cost, and opportunity cost. The longer it takes to make a decision, the greater the cost. Optimization problems exist across every industry \cite{Yarkoni2022}. They are difficult to solve due to incomplete or uncertain data, difficulties in stating the problem or examining the value of proposed solutions, or because they are \emph{NP-hard} problems. Stemming from the P-versus-NP-problem \cite{Fortnow2013}, the time required to solve NP-hard problems is suspected to grow exponentially with the size of the problem. The problems are \textit{hard} because of the large and exponentially growing solution space (combinatorial explosion). As a result, automotive companies such as BMW \cite{BMW2021} and Volkswagen \cite{Yarkoni2021,Yarkoni2019} are already exploring the possibility of quantum-based optimization for manufacturing. Additionally, quantum methods have been applied to portfolio optimization \cite{Egger2020,Orus2019,Yalovetzky2021}, logistics \cite{Bentley2022}, supply chain management \cite{Weinberg2022}, and energy economy \cite{Mastroianni2022}.

The aim of this perspective is to critically examine the status and future of quantum computing for optimization, which we define as \emph{quantum optimization}, including some possible weaknesses. We intend to complement the \textit{Royal Society}'s recent special edition on quantum annealing \cite{Chakrabarti2023} and \textit{Applied Quantum Computing}'s excellent review which provides accessible explanations for the technicalities of quantum computing and their practical uses \cite{Cumming2022}. We also recommend Ref \cite{guenin2014gentle} as an introduction to optimization. The outline of this paper is as follows. We identify suitable optimization problems that can benefit from quantum methods (Section~\ref{sec:usecase}). Then we address the challenges in formulating problems for quantum computers and algorithms (Section~\ref{sec:model}) before surveying the available algorithms (Section \ref{sec:algo}). In Section~\ref{sec:hardware}, we assess the current quantum computing hardware and determine possible weaknesses using a demonstration benchmark. Finally we present our perspective on quantum optimization, focusing on its present and future status.

\section{Finding Suitable Use Cases}\label{sec:usecase}

Compared to classical computers, quantum computers cope well with large search spaces, complex problem structures and objective functions. For example, quantum annealing is used to solve optimization problems which can take many forms: unconstrained versus constrained; using differential equations as constraints; combinatorial and graph problems; supposedly simple linear problems. (For an overview of real-world applications, see \cite{Hauke2020,Yarkoni2022}.) Many publications in quantum optimization focus on science and engineering such as the use of quantum annealing in the drug discovery and development process \cite{McKinsey2021,BCG2019b,Paul2010}. Due to significant investment in R\&D \cite{Brown2022,McKinsey2021,Wouters2020}, there are strong incentives to try new computational methods like Google DeepMind's breakthrough AlphaFold platform \cite{AlQuraishi2021,Senior2020} for the pharmaceutical industry. Quantum computing can overcome AlphaFold's shortcomings \cite{Wong2022,Mulligan2019} and dramatically change both landscape and revenue in the next few decades \cite{BCG2021}.

While there have been substantial efforts in applying quantum methods to specific problems, less work has been done on finding systematic approaches to determine which problems are most promising \cite{Chancellor2020method}. Use cases for quantum optimization are generally NP-hard or worse and difficult to approximate (with a constant factor). However, some NP-hard problems are solvable or heuristics/approximation algorithms provide acceptable results, e.g., the knapsack problem. A good example of an NP-hard problem is the \emph{Maximum Weighted Independent Set (MWIS)} problem: Given a topology graph, we want to find a set of vertices with maximum weight such that no vertices of that set are connected by an edge \cite{Lovasz1994, Pardalos1994}. The problem cannot be approximated to a constant factor unless P = NP \cite{Arora2009} and approximation to a polynomial factor is NP-hard itself \cite{Bazgan2005}. MWIS has practical applications, for example in DNA sequencing \cite{Joseph1992} and quantum computing \cite{Ebadi2022}. We use MWIS as our example problem for the rest of this paper.

\section{Modeling}\label{sec:model}

Solving real-world problems typically requires reformulating into a mathematical problem. Given a graph $G=G(V,E)$ with $m$ vertices, weight $a_i$ for each vertex $V$ and set of edges $E$, the classical formulation of the MWIS problem is
\begin{maxi*}[2]
{}{\sum_{i=1}^m a_i x_i}{}{}
\addConstraint{x_i+x_j}{\leq 1, }{\forall (i,j)\in E, }\addConstraint{x}{\in \{0,1\}^{m}.}{}
\end{maxi*}

For the quantum computing algorithms that we discuss in Section~\ref{sec:algo}, the optimization problem must be in the form of a \emph{quadratic unconstrained binary optimization problem (QUBO)}, $\min_{x\in\{0,1\}^m} x^\top Q x$. In quantum physics, the \emph{Hamiltonian} is used to calculate the system's minimum energy state. QUBOs are related to the \emph{Ising Hamiltonian} describing the energy of a solid in a ferromagnetic field using $n$ spins $s_i={\pm 1}$
\begin{equation*}\label{eq:qubo}
H(s_1,\ldots,s_n) = -\sum_{i<j} J_{ij} s_i s_j - \sum_{i=1}^n h_i s_i
\end{equation*}
with interaction strength $J_{ij}$ between two adjacent spins $i,j$ and spin-external magnetic field interaction $h_i$. Spins can be transformed to binary variables\footnote{See \cite{Zaman2021} for a QUBO to Ising mapping Python package.} via $x_i=(s_i+1)/2$. The general method for transforming problems to QUBO involves:
\begin{enumerate}
    \item Introducing slack variables \cite{Boyd2004} to transform inequality constraints to equality constraints.
    \item Transforming equality constraints to quadratic penalty terms.
    \item For integer and continuous variables, defining a discrete set of possible values and applying a binary encoding of these values.
\end{enumerate}

\subsection{Modelling Constraints}
While the procedure above works to transform any linear or quadratic problem, it has its flaws. In MWIS for example, we must introduce a binary slack variable $y_e$ and add penalty term $\left(x_i+x_j-y_e-1\right)^2$ to the objective function for every edge $e\in E$. As each quadratic term in the QUBO's objective function requires a connection between the qubits corresponding to the variables, the number of connections quickly exceeds what is available in current NISQ devices. For problems with integer and/or continuous variables the problem is exacerbated due to binary encoding. More sophisticated techniques have been presented by Glover et al. \cite{Glover2019, Glover2020} and Lucas \cite{Lucas2014} including QUBO formulations for Karp's 21 NP-hard problems \cite{Karp1972}. Advanced modeling has been applied to real-world problems \cite{Halffmann2022}. 

A better formulation for the MWIS is
\begin{maxi*}[2]
{}{\sum_{i=1}^m a_i x_i - p\sum_{(i,j)\in E} x_i x_j}{}{}
\addConstraint{x}{\in \{0,1\}^{m}}{}
\end{maxi*}
where we add no additional variables\footnote{For Ising model, set $n=m$ and $s_i=2x_i-1$.} and have less connectivity. 
The challenging task of choosing the right scalar penalty factor $p$ relies on domain knowledge and experience. We can also obtain decent estimates from the constraints (e.g. setting $p\coloneqq \max_i a_i\cdot \max_i deg(i)+1$, where $deg(i)$ is the number of edges vertex $i$ is connected with). High penalty factors cause the original objective function to lose its significance, making it hard to distinguish between optimal and feasible solutions, particularly on NISQ devices. Feasible solutions are disregarded when penalties are too small. To our knowledge, no formal techniques exist although simple machine learning approaches could improve the formulation quality.

\subsection{Modelling Variables}

\textit{Binary encoding} \cite{Lucas2014} is widely used for higher-than-binary model variables. The number of possible values a variable can attain is reduced to $M$ which then can be encoded using $\lfloor \log(M)\rfloor +1$ qubits. It is resource intensive, thus we highlight two alternative approaches. 

Domain-wall is a recently proposed method for encoding discrete variables \cite{chancellor2019a}, dramatically improving both gate-model variational algorithms \cite{plewa2021variational} and annealers \cite{Abel20a,chen21a}. In this method, arbitrary two-variable interactions between discrete variables of size $m$ can be represented in an encoding which requires $m-1$ qubits per variable, one less than the more traditional one-hot encoding method. Importantly, arbitrary pairwise interactions between discrete variables can be mapped to quadratic interactions between qubits. It is suitable for devices with limited connectivity \cite{chancellor2019a} and uses the minimum number of qubits where such a guarantee is possible \cite{berwald2021understanding}. It has been explored for networking applications has \cite{Chen2022a}.

A recent work \cite{Bermejo2022} deviates from the standard QUBO approach by considering continuous variables without binary encoding. The qubit state can be given as a position on or in the Bloch sphere for pure and mixed states respectively, so they make use of the 3D coordinates given by two angles and the radius. They can encode two (three) continuous variables into each pure- (mixed-)state qubit.

\section{Algorithms}\label{sec:algo}

Here we present two popular methods for solving optimization problems: variational quantum algorithms and quantum annealing. Further methods exist like Grover adaptive search \cite{Gilliam2021} and hybrid algorithms \cite{Ajagekar2020,Callison2022} but these are beyond the scope of our work. For an exhaustive list of quantum algorithms, we recommend the Quantum Algorithm Zoo \cite{Zoo}.

\subsection{Variational quantum algorithms}

Variational quantum algorithms (VQAs) provide a general framework for solving many problems such as big data analysis and simulations for quantum chemistry and materials science \cite{Kandala2017,Kokail2019}. Their adaptive nature is due to the variational process: the optimization procedure varies the algorithm on-the-fly, similar to machine learning \cite{Cerezo2021}. The problem is encoded into a parametrized cost function that defines a hypersurface. The quantum circuit navigates this surface to estimate the global minima which correspond to solutions of the problem. The result is passed back to the classical computer to adjust the VQA parameters. Then this is returned to the quantum computer to repeat the process.

VQAs require an ansatz whose form dictates what the variational parameters are and how they can be trained to minimize the cost function. The ansatz structure depends on the problem at hand. One major bottleneck is the barren plateau phenomenon, where changes in the classical parameters have little effect on the optimality of the variational state. The presence of barren plateaus is tied to the expressivity of the variational ansatz \cite{Holmes2022a}. Highly expressive ansatzes tend to cause the algorithms to become effectively lost in a vast space of highly sub-optimal solutions. This leads to a tradeoff: if the ansatz is not expressive enough, it may not be able to effectively solve the problem; if too expressive, then the algorithm cannot be efficiently trained \cite{Callison2022,Cerezo2021}.

There are two major VQAs: 
\begin{enumerate}
\item The computationally universal quantum approximate optimization algorithm (QAOA) \cite{Hadfield2019,Lloyd2018a,Morales2020,Zhou2020}, originally introduced to solve combinatorial optimization problems \cite{Farhi2014,Farhi2014b}.

\item Variational quantum eigensolvers (VQEs) \cite{Cerezo2021,Tilly2022} are the best-known application of VQAs and were originally used to find the ground state energy of molecules \cite{AspuruGuzik2005,McClean2016,Wecker2015}.

\end{enumerate}

\subsection{Quantum annealing and D-Wave} \label{sec:quantumannealing}

\begin{figure*}[ht!]
\centering
\begin{tikzpicture}[font=\sffamily,thick]

\draw[fill=OliveGreen!25,draw=OliveGreen,fill opacity=0.4]  (12,1.5) rectangle ++(-4,1);
\draw[fill=NavyBlue!25,draw=NavyBlue,fill opacity=0.4]  (12,0.5) rectangle ++(-4,1);
\draw[fill=OliveGreen!25,draw=OliveGreen,fill opacity=0.4] (17,0.5) -- (13,0.5) -- plot[smooth, tension=.7] coordinates {(13,1) (13.3,1.25) (13.75,1.9) (14.5,1.3) (15.95,1.45) (16.4,2.25) (16.7,1.55) (17,1.4)} -- (17,0.5);
\draw[fill=NavyBlue!25,draw=NavyBlue,fill opacity=0.4] (13,2.5) -- (13,0.5) -- (17,0.5) -- plot[smooth, tension=.7] coordinates {(17,2) (16.7,1.95) (16.25,1.5) (15.45,3) (14.9,1.7) (14.5,1.65) (13.7,1) (13.3,2.6) (13,2.5)};

\draw[fill=OliveGreen!25,draw=OliveGreen,fill opacity=0.5] (22,0.5) -- (18,0.5) -- plot[smooth, tension=.7] coordinates {(18,1) (18.6,1.4) (18.85,2.15) (19.1,1.45) (19.6,1.15) (20.3,1.05)   (22,1)} -- (22,0.5);
\draw[fill=NavyBlue!25,draw=NavyBlue,fill opacity=0.5] (18,2.4) -- (18,0.5) -- (22,0.5) -- plot[smooth, tension=.7] coordinates {(22,2.25) (21.7,1.75) (20.95,1.55) (20.35,1.95) (19.9,1.4) (19.5,1.8) (19.25,1.55) (18.8,0.9) (18.45,1.55) (18.2,3.15) (18,2.5)};

\draw (8,3.5) -- (8,0.5) -- (12,0.5) node (v1) {} -- (12,3.5);
\node at (10,0) {\normalsize configuration};
\node at (10,4) {\normalsize $t=0$};
\node[rotate=90,color=OliveGreen] at (7.5,2) {\textbf{probability}};
\draw (13,3.5) -- (13,0.5) -- (17,0.5) -- (17,3.5);
\node at (15,0) {\normalsize configuration};
\node at (15,4) {\normalsize $0<t<T$};
\node at (13.85,2.15) {$\vert\psi_1\rangle$};
\node at (16.5,2.6) {\small $\vert\psi_2\rangle$};
\draw (18,3.5) -- (18,0.5) -- (22,0.5) -- (22,3.5);
\node at (20,0) {\normalsize configuration};
\node at (20,4) {$t=T$};
\node[rotate=90,color=NavyBlue] at (22.5,2) {\normalsize \textbf{energy}};
\node at (18.95,2.5) {{\small $\vert\psi_1\rangle$}};

\tikzstyle{cc} = [align=left,font=\sffamily\footnotesize]

\node[cc] at (15,16.5) {\textbf{formulate} problem \\ into Ising or QUBO \\ model};
\node[cc] at (15,14) {\textbf{represent} model \\ as a qubit graph \\ network};
\node[cc] at (15,11.5) {\textbf{embed} graph \\ in hardware \\ structure};
\node[cc] at (15,9) {\textbf{specify} parameters, \\ incl. annealing time \\ and sampling \\ strategy};
\node[cc] at (15,6.5) {{\normalsize \textbf{perform}} \\ {\normalsize quantum} \\ {\normalsize annealing}};
\node[cc] at (21,6.5) {\textbf{unpack} results of \\ hardware structure \\ to get output};
\node[cc,draw=BrickRed] at (18.75,7.85) {\textbf{resample} when necessary};

\draw (13.5,17.5) rectangle (16.5,15.5);
\draw (13.5,15) rectangle (16.5,13);
\draw (13.5,12.5) rectangle (16.5,10.5);
\draw (13.5,10) rectangle (16.5,8);
\draw[line width=2pt,draw=BlueViolet]  (13.5,7.5) rectangle (16.5,5.5);
\draw (19.5,7.5) rectangle (22.5,5.5);

\draw[arrows = {-Stealth[inset=0pt, angle=90:20pt]}, line width=2pt] (15,15.5) -- (15,15);
\draw[arrows = {-Stealth[inset=0pt, angle=90:20pt]}, line width=2pt] (15,13) -- (15,12.5);
\draw[arrows = {-Stealth[inset=0pt, angle=90:20pt]}, line width=2pt] (15,10.5) -- (15,10);
\draw[arrows = {-Stealth[inset=0pt, angle=90:20pt]}, line width=2pt] (15,8) -- (15,7.5);
\draw[-latex, line width=4pt] (16.5,6.5) -- (19.5,6.5);
\draw[-latex, line width=4pt, color=BrickRed] (19.25,7.25) -- (16.5,7.25);
\draw[draw=BlueViolet, rounded corners, line width=2pt]  (7,4.5) rectangle (23,-0.5);

\draw[draw=BlueViolet, line width=2pt] (15,5) ellipse (.7 and 0.5);
\draw[arrows = {-Stealth[inset=0pt, angle=45:10pt]}, draw=BlueViolet, line width=4pt] (15.65,5.1) -- (15.7,4.95);
\draw[arrows = {-Stealth[inset=0pt, angle=45:10pt]}, draw=BlueViolet, line width=4pt] (14.4,4.75) -- (14.25,5);
\end{tikzpicture}
\caption{Quantum annealing process, including time evolution from $H(0)=H_0$ to $H(T)=H_p$.}
\label{fig:annealing}
\end{figure*}
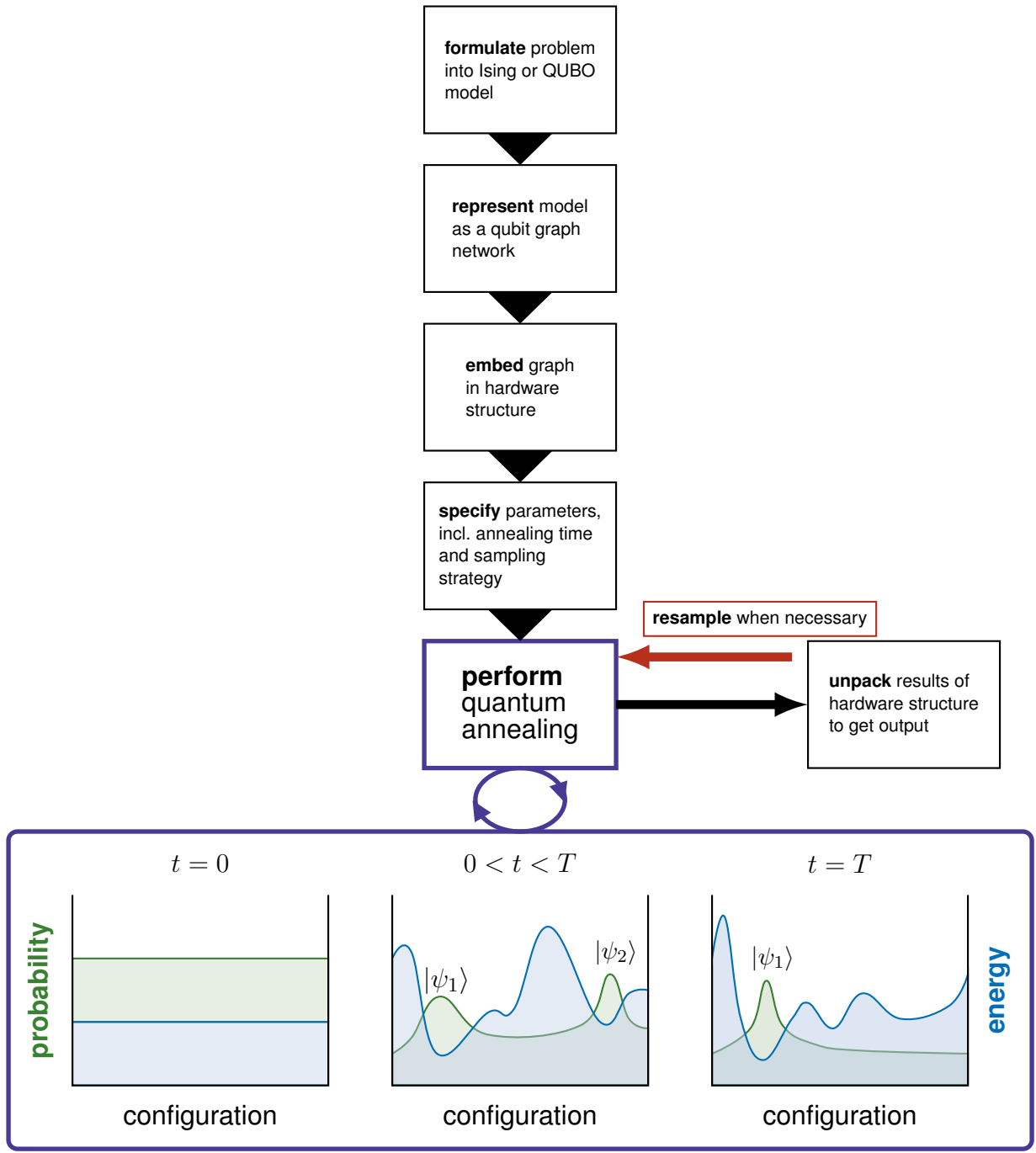

Quantum annealing \cite{Hauke2020,Kadowaki1998,Yarkoni2022} is a form of quantum computing based on continuous time evolution. Typically it operates outside the regime of adiabatic quantum computing which gives theoretical guarantees of performance \cite{Albash2018,Farhi2000}. The Hamiltonian takes the general form
\begin{equation}
H(t) = H_0(t) + H_p(t) = \left(1-\frac{t}{T}\right)H_0 + \frac{t}{T}H_p.
\end{equation}
Initially ($t=0$), quantum annealing begins in a prepared state of Hamiltonian $H_0$ with uniform probability (Fig. \ref{fig:annealing}). During the system evolution ($0<t<T$), probabilities are gradually driven toward the global minimum (near-optimal solution) by final time $T$ and the system is in the eigenstate of problem's Hamiltonian $H_p$. 

Quantum annealing can be done on D-Wave computers. Evolution is governed by the transverse field Ising Hamiltonian,
\begin{align}
H(t) = A(t) \underbrace{\sum_n \sigma^x_n}_{H_0} + 
B(t) \underbrace{\sum_{n,m} 
\left( 
h_n \sigma^z_n + 
J_{nm} \sigma^z_n \sigma^z_m 
\right)}_{H_p}
\end{align}
with Pauli-$x$ and -$z$ matrices $\sigma^{x,z}$, symmetric interaction strength $J_{nm}=J_{mn}$ of qubits $n$ and $m$, and on-site energy $h_n$.

\section{Hardware and Benchmarking}\label{sec:hardware}

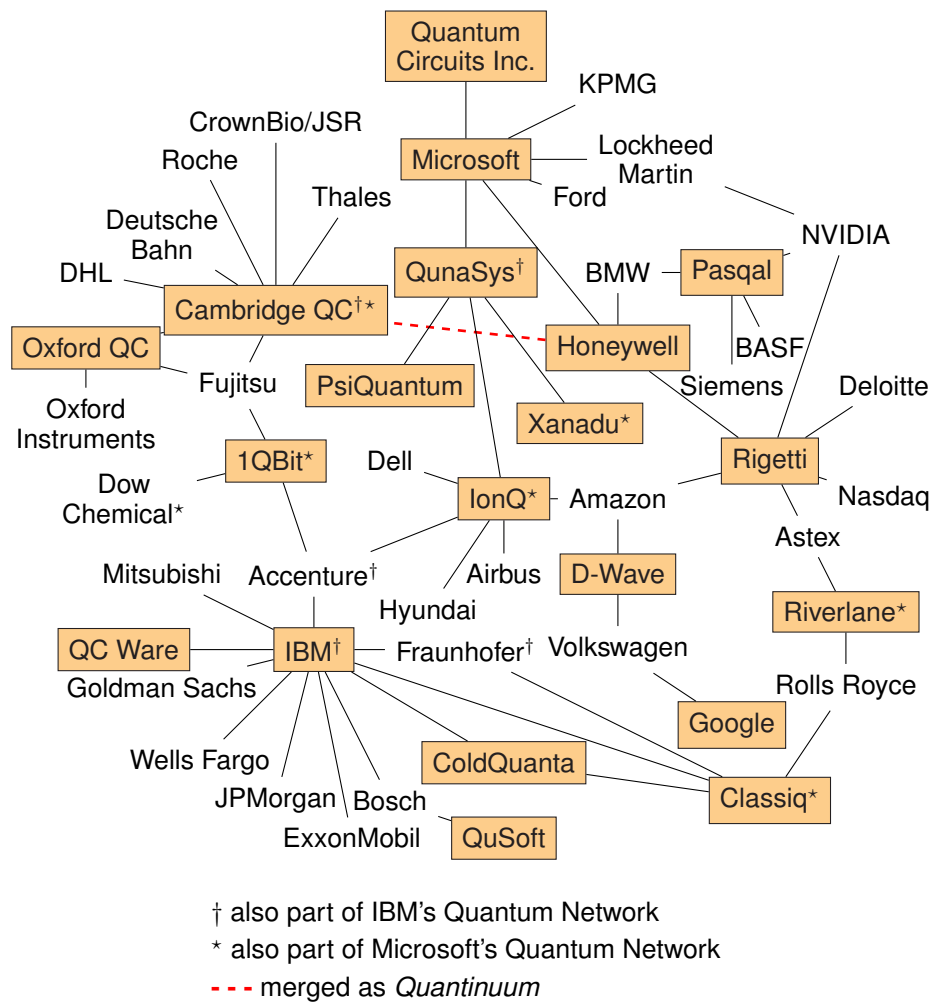
\begin{figure*}[ht!]
\centering
\begin{tikzpicture} 
\tikzstyle{every node}=[font=\tiny\sffamily,align=center]
\tikzstyle{su} = [draw=black, fill=YellowOrange!50, text=black, align=center]
\node [su] (v1) at (4.5,-1.5) {Honeywell};
\node [su] (v3) at (6,-0.5) {Pasqal};
\node [su] (v4) at (0.5,-5.5) {IBM$^\dagger$};
\node [su] (v8) at (0,-1) {Cambridge QC$^\dagger$$^\star$};
\node [su] (v10) at (6.5,-3) {Rigetti};
\node [su] (v14) at (-2.5,-1.5) {Oxford QC};
\node [su] (v18) at (6.5,-7.5) {Classiq$^\star$};
\node [su] (v20) at (7.5,-5) {Riverlane$^\star$};
\node [su] (v22) at (4.5,-4.5) {D-Wave};
\node [su] (v25) at (2.5,1) {Microsoft};
\node [su] (v26) at (0,-3) {1QBit$^\star$};
\node [su] (v29) at (3,-3.5) {IonQ$^\star$};
\node [su] (v36) at (2.5,2.5) {Quantum \\ Circuits Inc.};
\node [su] (v40) at (3,-8) {QuSoft};
\node [su] (v41) at (6,-6.5) {Google};
\node [su] (v44) at (3,-7) {ColdQuanta};
\node [su] (v45) at (-2,-5.5) {QC Ware};
\node [su] (v48) at (2.5,-0.5) {QunaSys$^\dagger$};
\node [su] (v49) at (1.5,-2) {PsiQuantum};
\node [su] (v50) at (4,-2.5) {Xanadu$^\star$};
\node (v2) at (4.5,-0.5) {BMW};
\node (v5) at (1,-8) {ExxonMobil};
\node (v6) at (0,-7.5) {JPMorgan};
\node (v7) at (6,-2) {Siemens};
\node (v9) at (-1.5,0) {Deutsche \\ Bahn};
\node (v11) at (7.5,0) {NVIDIA};
\node (v12) at (1.5,-7.5) {Bosch};
\node (v13) at (-1.5,-6) {Goldman Sachs};
\node (v15) at (-2.5,-2.5) {Oxford \\ Instruments};
\node (v16) at (-0.5,-2) {Fujitsu};
\node (v17) at (8,-3.5) {Nasdaq};
\node (v19) at (7.5,-6) {Rolls Royce};
\node (v21) at (4.5,-5.5) {Volkswagen};
\node (v23) at (-1,1) {Roche};
\node (v24) at (-2.5,-0.5) {DHL};
\node (v27) at (-2,-3.5) {Dow \\ Chemical$^\star$};
\node (v28) at (5,1) {Lockheed \\ Martin};
\node (v30) at (3,-4.5) {Airbus};
\node (v32) at (2,-5) {Hyundai};
\node (v31) at (1.5,-3) {Dell};
\node (v33) at (0.5,-4.5) {Accenture$^\dagger$};
\node (v34) at (4.5,-3.5) {Amazon};
\node (v35) at (0,1.5) {CrownBio/JSR};
\node (v37) at (1,0.5) {Thales};
\node (v38) at (8,-2) {Deloitte};
\node (v39) at (6.5,-1.5) {BASF};
\node (v42) at (2.5,-5.5) {Fraunhofer$^\dagger$};
\node (v43) at (7,-4) {Astex};
\node (v46) at (-1,-7) {Wells Fargo};
\node (v47) at (-1.5,-4.5) {Mitsubishi};
\node (v51) at (4.5,2) {KPMG};
\node (v52) at (4,0.5) {Ford};

\draw (v1) -- (v2); \draw[dashed,line width=1pt,draw=red] (v1) -- (v8); \draw (v1) -- (v10); \draw (v1) -- (v25); 
\draw (v2) -- (v3); 
\draw (v3) -- (v7); \draw (v3) -- (v11); \draw (v3) -- (v39); 
\draw (v4) -- (v5); \draw (v4) -- (v6); \draw (v4) -- (v12); \draw (v4) -- (v13); \draw (v4) -- (v18); \draw (v4) -- (v33); \draw (v4) -- (v42); \draw (v4) -- (v44); \draw (v4) -- (v45); \draw (v4) -- (v46); \draw (v4) -- (v47); 
\draw (v8) -- (v9); \draw (v8) -- (v14); \draw (v8) -- (v16); \draw (v8) -- (v23); \draw (v8) -- (v24); \draw (v8) -- (v35); \draw (v8) -- (v37); 
\draw (v10) -- (v11); \draw (v10) -- (v17); \draw (v10) -- (v34); \draw (v10) -- (v38); \draw (v10) -- (v43); 
\draw (v11) -- (v28); 
\draw (v12) -- (v40); 
\draw (v14) -- (v15); \draw (v14) -- (v16); 
\draw (v16) -- (v26); 
\draw (v18) -- (v19); \draw (v18) -- (v44); \draw (v18) -- (v42); 
\draw (v19) -- (v20); 
\draw (v20) -- (v43); 
\draw (v21) -- (v22); \draw (v21) -- (v41); \draw (v22) -- (v34); 
\draw (v25) -- (v28); \draw (v25) -- (v36); \draw (v25) -- (v48); \draw (v25) -- (v51); \draw (v25) -- (v52); \draw (v26) -- (v33); \draw (v26) -- (v27); 
\draw (v29) -- (v30); \draw (v29) -- (v31); \draw (v29) -- (v32); \draw (v29) -- (v33); \draw (v29) -- (v34); \draw (v29) -- (v48); 
\draw (v48) -- (v49); \draw (v48) -- (v50); 

\node[right] at (-1,-9) {$\dagger$ also part of IBM's Quantum Network};
\node[right] at (-1,-9.5) {$^\star$ also part of Microsoft's Quantum Network};
\node[right] at (-1,-10) {\textbf{\textcolor{red}{- - -}} merged as \emph{Quantinuum}};
\end{tikzpicture}
\caption{Quantum computing commercial landscape (selected companies only). Academic institutions and government organizations not shown. Details taken from press releases and publicly disclosed interactions with end users.}
\label{fig:commercial_landscape}
\end{figure*}

There are numerous quantum computer manufacturing companies ranging from small startups to large well-established companies \cite{MacQuarrie2020}. We show examples of collaborations in Fig. \ref{fig:commercial_landscape}. A leading platform for solving optimization problems is D-Wave Systems which are based on a superconducting flux-qubit design \cite{Bunyk2014}. Other platforms such as IBM Q use superconducting electronics (transmons). For brevity, we focus on the most mature platforms with over 100 qubits: D-Wave and IBM. While D-Wave only applies quantum annealing (Section~\ref{sec:quantumannealing}), IBM's gate-based universal quantum computer allows various algorithms. D-Wave has been selling quantum computers since 2011 \cite{Nimbe2021} and has become an important player in the quantum computer industry \cite{BCG2018}. Both D-Wave and IBM's methods involve time-evolving quantum systems \cite{crosson2021prospects}, and the companies themselves have ambitious plans to shape the future of quantum computing\footnote{IBM Roadmap \url{https://www.ibm.com/quantum/roadmap},\\ D-Wave roadmap \url{https://www.dwavesys.com/media/xvjpraig/clarity-roadmap_digital_v2.pdf}.}.

\begin{figure}[ht!]
\centering
\includegraphics[width=\linewidth]{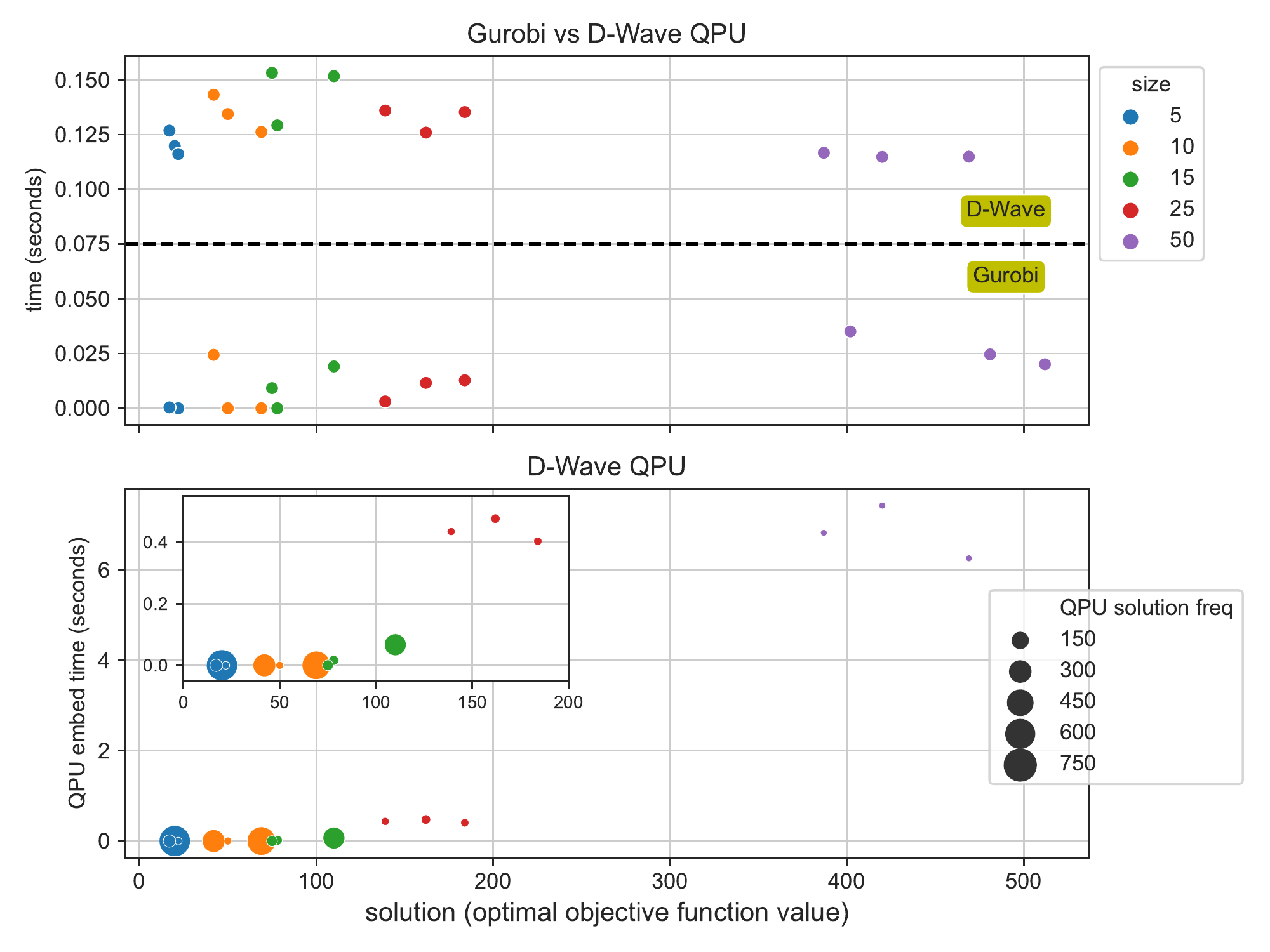}
\caption{Benchmarking results for MWIS problem. For Gurobi, we show the optimal objective function value and total running time in seconds. For D-Wave, we show the embedding time and QPU access time required to solve the problem (both in seconds), plus the best solution found and its frequency in 1,000 shots. See data table in supplementary information.}
\label{fig:benchmark}
\end{figure}

To compare quantum and classical methods, we perform a demonstration benchmark for the MWIS problem. We consider randomly generated graphs with different numbers of vertices (size $k$) and randomly selected vertex weights between $1$ and $2k+1$. We compare D-Wave Quantum Annealer's\footnote{Version Advantage System 5.2 with over 5,000 qubits, 1,000 shots.} results to Gurobi 10.0 \cite{gurobi} on classical hardware\footnote{Intel(R) Core(TM) i7-8665U CPU  1.90GHz, 16 GB RAM, 250GB SSD using Windows 10 and Python 3.9} (Fig. \ref{fig:benchmark}). Gurobi outperforms D-Wave, but the results up to size 25 are promising (Fig. \ref{fig:benchmark}) because QPU embedding time is a similar order of magnitude to a state-of-the-art commercial solver. Increasing the number of vertices decreases solution quality and probability. Additionally, the embedding time dominates the total time required to solve the problem. Instances up to size 25 have also been tested on IBM via QAOA with warm start \cite{Egger2021}. IBM Qiskit\footnote{Version 0.39.4 QASM with state vector simulator} can calculate optimal solutions up to size 10 but fails for size 25 and above. Using the Falcon r6 device\footnote{In Ehningen, Germany.} with 27 qubits, only a graph of size 5 occasionally gave the correct solution. Note that Gurobi can certify the solution correctness by using (for example) duality, and Gurobi confirmed optimality for all results. D-Wave do not certify their results.

To summarize, these results suggest that the classical approach is more efficient and that there is no practical quantum advantage for small numbers of graph vertices. Hence we may ask whether this is the best benchmarking strategy. Better and fairer strategies exist \cite{Katzgraber2014}, especially for D-Wave, and for analyzing how results will scale \cite{Weinberg2020}. However this is beyond the scope of this work where our benchmark is intended as a proof-of-concept. Only with more thorough benchmarking can we accurately assess the capabilities and speed-ups of quantum computers \cite{Ronnow2014}.

\section{Discussion and Outlook}\label{sec:outlook}

In this work, we surveyed quantum methods for solving optimization problems. Our demonstration benchmarking compared quantum and classical methods for solving the MWIS problem. Now we discuss several key points and our perspectives on quantum optimization.

Finding a good problem formulation has a key impact on the solution quality. Historically, optimization practitioners have preferred constrained linear representations of problems. Quantum computing presents a different paradigm where the fundamental objects for optimization are quadratic expressions. Here we should minimize the number of qubits and connections. Mapping the Hamiltonian onto quantum hardware architecture is challenging but can be significantly improved via applying specialized graph optimization algorithms and machine learning methods. We stress that the transformation to QUBO is not inevitable. For example, Grover's adaptive search \cite{Bulger2003,Gilliam2021} can take any function at the cost of deeper circuits. While this is promising, practical advantage (compared to say, quantum annealers) is not definitive.

Practical application of quantum computing requires further progress in hardware. Measuring hardware quality and that of formulations and algorithms likewise is a research task on its own. Here, we need better evaluation and benchmarking methods, particularly for practical use cases where contested assertions like quantum supremacy\footnote{While there is unfortunately no unambiguous way to refer to the concept, we acknowledge that aspects of this phrase are problematic \cite{Palacios-Berraquero2019}.} do not play a role. For example, Tang proposed that we should \enquote{dequantize} quantum machine learning algorithms to check whether any quantum speedup exists \cite{Tang2022}.
    
Quantum computing R\&D is a highly interdisciplinary process. Some parts of the research community believe that quantum hardware must be built first before we develop algorithms for applications. However to leverage the expertise of all participants, we should approach the entire quantum computing stack simultaneously, allowing cross-pollination of ideas across the stack \cite{Fu2016,vanMeter2013}. Engineers can build appropriate quantum devices when they know the problem statement and how best to solve it, whereas theorists can design algorithms or software once they understand the hardware capabilities and limitations.

Similarly, quantum optimization algorithms will benefit from classical approaches. Quantum algorithms often borrow ideas from classical methods (cf. quantum annealing and classical simulated annealing \cite{Bertsimas1993,Kirkpatrick1983}). However, few established optimization techniques other than, say branch-and-bound \cite{Montanaro2020}, have been adapted for quantum computation. We are confident that combining with classical methods can improve the performance of quantum optimization algorithms. For example, decomposition techniques implemented in mixed-integer solvers can split the problem in two and assign each part to classical and quantum hardware \cite{Zhao2022}.

Not every hard problem will benefit from quantum algorithms. For many hard problems, classical solvers are so efficient that quantum methods do not provide appreciable speed-ups, e.g. for the knapsack problem and our MWIS example. For example it is naive to assume that quantum optimization can speed up problems in P, e.g., linear continuous problems. An exception may be that particular subroutines could be improved with quantum algorithms. Hence despite the excitement over quantum speed-up, quantum computing itself is not a revolution but more like a new numerical toolbox that uses different processing units. It is likely that many future high performance computers (or computer centers) will contain a stack of CPUs, GPUs and QPUs similar to the JUNIQ program at the J{\"u}lich Supercomputing Centre \cite{juniq2019launch}. Such a platform could involve one control unit that uses the CPUs, GPUs and QPUs interchangeably for each part of the computation. We believe  this is where quantum computing has the most potential for impact. 

Quantum computing and optimization are forecast to be competitive in under a decade \cite{BCG2019}. Now is the time to start planning for the future and hence maximize quantum computing's commercial impact \cite{BCG2018}. The best way to turn potential obstacles into opportunities is through collaboration between quantum theorists, computer scientists, engineers, end-users, and beyond. Then we can truly benefit from quantum optimization.


\section*{Conflict of Interest Statement}

R. A. and P. H. declare that the research was conducted in the absence of any commercial or financial relationships that could be construed as a potential conflict of interest. Although they did not directly fund this work N. C. was employed part time as a consultant by Quantum Computing Inc.~at the time this manuscript was written.

\section*{Author Contributions}
$\dagger$ All authors contributed equally to this work and share first authorship.

\section*{Contribution to the Field Statement}
Quantum computing has attracted much interest from academia, politics and industry. This is due to the high expectations but also rapid advances in the last five years. Numerous application areas exist for quantum computing, particularly optimization. In this perspective, we investigate a state-of-the-art quantum method for solving optimization problems using \enquote{quantum optimization}. Our method follows the traditional modeling cycle: we observe which use-cases are suitable for quantum optimization. Problem formulations for quantum algorithms and hardware require new modeling paradigms. We describe several quantum computing algorithms and give an overview on quantum hardware. This is accompanied by a demonstration benchmarking where we conclude that classical approaches outperform quantum optimization in some cases, even though the results are competitive. Finally, we assess the future of quantum optimization and key points, such as combining quantum computing with established optimization techniques to inspire possibilities for future research.

\section*{Funding}

R.~A.~acknowledges support from UK Quantum Technology Hub in Computing and Simulation (grant EP/T001062/1) and UKRI (grant EP/W00772X/2). N.~C.~was supported by UKRI grants EP/W00772X/1
 and EP/T026715/1. P.~H.~is funded by the German Federal Ministry for Economic Affairs and Climate Action (grant no. 03EI1025A).

\section*{Acknowledgments}

The authors are grateful for discussions with V. Kendon. The authors thank P. Holzer and M. Trebing for generating the results on IBM and D-Wave.

\bibliographystyle{Frontiers-Harvard} 
\bibliography{frontiers}

\end{document}